\documentclass{article}

\usepackage{amssymb}

\usepackage[figuresright]{rotating}
\usepackage{amsmath}
\usepackage{xspace}

\newcommand{\ccbar}{{$\text{c} \overline{\text{c}}$ }\xspace}
\newcommand{\accbar}{{$\langle \text{c} \overline{\text{c}}\rangle$ }\xspace}
\newcommand{\AGeVc}{\ensuremath{A\,\mbox{Ge\kern-0.1em V}\!/\kern-0.08em c}\xspace}
\newcommand{\AGeV}{\ensuremath{A\,\mbox{Ge\kern-0.1em V}}\xspace}

\newcommand{\pip}{\ensuremath{\pi^+}}
\newcommand{\km}{\ensuremath{\textup{K}^-}\xspace}

\newcommand{\Dzero}{\ensuremath{\textup{D}^0}\xspace}
\newcommand{\Dzerobar}{\ensuremath{\overline{\textup{D}^0}}\xspace}
\newcommand{\Dplus}{\ensuremath{\textup{D}^+}\xspace}
\newcommand{\Dminus}{\ensuremath{\textup{D}^-}\xspace}

\title{Open charm measurements in NA61/SHINE at CERN SPS}

\author{Pawel Staszel for the NA61/SHINE Collaboration \\
Jagiellonian University Institute of Physics, \\ ul. Lojasiewicza 11, 30-348 Krakow, PL}

\begin{document}

\maketitle





\begin{abstract}

The measurements of open charm production was proposed as an important tool to 
investigate the properties of hot and dense matter formed in nucleus-nucleus collisions 
as well as to provide the means for model independent interpretation of the existing data on $\text{J}/\psi$ suppression. 
Recently, the experimental setup of the NA61/SHINE 
experiment was supplemented with a Vertex Detector which was motivated 
by the importance and the possibility of the first direct 
measurements of open charm meson production in heavy ion collisions at SPS energies. 
First test data taken in December 2016 on Pb+Pb collisions at 150$A$ GeV/$c$ allowed to validate the general concept 
of D$^0$ meson detection via it's D$^0~\rightarrow~\pi^{+}~+~K^{-}$ decay channel and delivered a first 
indication of open charm production.


The physics motivation of open charm measurements at SPS energies, pilot results on open charm 
production, and finally, the future plans of open charm measurements in the NA61/SHINE experiment after LS2 are presented. 

\end{abstract}




\section{Introduction}
\label{sec1}

The SPS Heavy Ion and Neutrino Experiment (NA61/SHINE) \cite{na61} is a fixed-target experiment located 
at the CERN Super Proton Synchrotron (SPS). The NA61/SHINE detector is optimized 
to study hadron production in hadron-proton, hadron-nucleus and nucleus-nucleus 
collisions.  
The strong interaction research program of NA61/SHINE is dedicated to the study of the properties of the onset of 
deconfinement and the search for the critical point of strongly interacting matter. 
These goals are being pursued by investigating p+p, p+A and A+A collisions at different beam momenta 
from 13A to 150A GeV/c.
In 2016 NA61/SHINE was upgraded with the Small Acceptance Vertex Detector (SAVD) based on 
MIMOSA-26AHR sensors developed in IPHC Strasbourg.  
Construction of this device was mostly motivated by the importance and the possibility of the first 
direct measurements of open charm meson production in heavy ion collisions at SPS energies.  
Precise measurements of charm hadron production by NA61/SHINE are expected to be performed in 2022--2024. 
The related preparations have started already.  

\section{Physics motivation for open charm measurements}
\label{sec2}

\begin{figure}[!htb]
	\centering
	\includegraphics[width=0.8\textwidth]{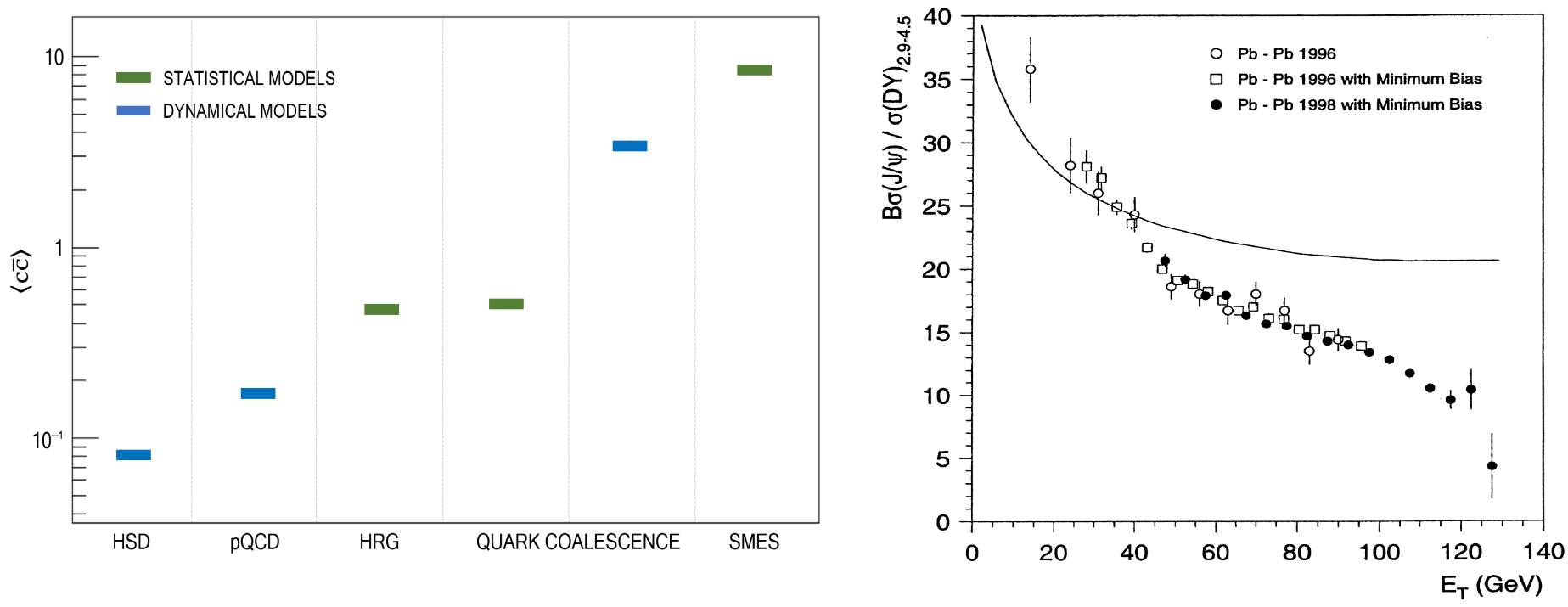}
        \vspace{-0.25cm}
	\caption{(\textit{Left:})
		Mean multiplicity of charm quark pairs produced in the full phase space in central Pb+Pb 
                collisions at 158\AGeVc calculated with dynamical models (blue bars): HSD~\cite{Linnyk:2008hp,TSong}, 
                pQCD--inspired~\cite{Gavai:1994gb,BraunMunzinger:2000px},  
		and Dynamical Quark Coalescence~\cite{Levai:2000ne}, as well as statistical 
                models (green bars): HRG~\cite{Kostyuk:2001zd}, Statistical Quark Coalescence~\cite{Kostyuk:2001zd}, 
                and SMES~\cite{Gazdzicki:1998vd}.
                (\textit{Right:})The ratio of $\sigma_{\text{J}/\psi}/\sigma_{\mathrm{DY}}$ as a function of 
                transverse energy (a measure of collision violence or centrality)
                in Pb+Pb collisions at 158\AGeV measured by NA50. 
                The curve represents the $ \text{J}/\psi $ suppression due to ordinary 
                nuclear absorption~\cite{Abreu:2000ni}.}
	\label{fig:models}
\end{figure}

One of the important aspects of relativistic heavy-ion collisions is the mechanism of charm
production. Several models were developed to describe charm production. Some of them
are based on dynamical and others on statistical approaches. The estimates from these
models for the average number of produced $\text{c}$  and $\overline{\text{c}}$ pairs 
(\accbar) in central Pb+Pb collisions at 158\AGeVc differ by up to a factor of 50 \cite{ana1,ana2} ss illustrated in
Fig.~\ref{fig:models} (\textit{left}).
Therefore, obtaining precise data on \accbar will allow to
distinguish between theoretical predictions and learn about the charm quark and hadron production
mechanism. A good estimate of  \accbar can be obtained by measuring the yields of \Dzero , \Dplus and their antiparticles
because these mesons carry about 85\% of the total produced charm \cite{ana3}.

Charm mesons are of special interest in the context of the phase transition between confined
hadronic matter and the quark gluon plasma (QGP).
The \ccbar pairs produced in the collisions are converted into open charm mesons and charmonia
($\text{J}/\psi$ mesons and the excited states). The production of charm is expected to be different in confined
and deconfined matter. This is caused by different properties of charm carriers in these phases. In
confined matter the lightest charm carriers are D mesons, whereas in deconfined matter the lightest
carriers are charm quarks. Production of a DD
pair (2m$_{D} = 3.7$ GeV) requires an energy about 1 GeV higher than production of a \ccbar
pair (2m$_{c} = 2.6$ GeV). The effective number of degrees of freedom of
charm hadrons and charm quarks is similar \cite{ana7}. Thus, in the statistical approach more abundant charm production is expected
in deconfined than in confined matter. Consequently, in analogy to strangeness production~\cite{ana2,ana8}, a change of
collision energy dependence of \accbar may be a signal of the onset of deconfinement.





Figure~\ref{fig:models} (\textit{right})  shows results on  $ \langle \text{J}/\psi \rangle $ production normalized to 
the mean multiplicity of Drell-Yan pairs in Pb+Pb collisions at the top SPS 
energy obtained by the NA50 collaboration. 
The solid line shows a model prediction for normal nuclear absorption of $\text{J}/\psi$ in the medium. 
NA50 observed that $\text{J}/\psi$ production is
consistent with normal nuclear matter absorption for peripheral collisions and is suppressed for more central collisions. 
This so called anomalous suppression was attributed to the  $\text{J}/\psi$ dissociation effect in the deconfined medium. 
However, the above result is based on the assumption that \accbar  $\sim \langle DY \rangle $ which may be incorrect 
due to several effects, such as shadowing or parton energy loss ~\cite{Satz:2014usa}.
Thus the effect of the medium on \ccbar binding can only be quantitatively determined by comparing 
the ratio of $\langle \text{J}/\psi \rangle$ to \accbar in nucleus-nucleus to that in proton-proton reactions. 
In Pb+Pb collisions the onset of color screening should already be seen in the centrality dependence 
of the $\langle \text{J}/\psi \rangle$ to \accbar ratio.
This clearly shows the need for large statistic data on \accbar. 

\section{Performance of SAVD}
\label{sec3}

The SAVD was built using sixteen CMOS MIMOSA-26 sensors~\cite{mimosa26}. The basic sensor properties are:
$18.4 \times 18.4~\mu$m$^2$  pixels, 115~$\mu$s time resolution,
$ 10 \times$ 20~mm$^2$  surface, 0.66 MPixel, 50~$\mu$m thick.	
The estimated material budget per layer, including the mechanical support, is 0.3\% of a radiation length.
The sensors were glued to eight ALICE ITS ladders~\cite{Abelev:1625842}, 
which were mounted on two horizontally movable arms and spaced by 5~cm 
along the \textit{z} (beam) direction. The detector box was filled with He (to reduce beam-gas interactions) 
and contained an integrated target holder to avoid unwanted material 
and multiple Coulomb scattering between target and detector. More details related to the SAVD project can be found in
\cite{deveaux}.

\begin{figure}[!htb]
	\centering
	\includegraphics[width=0.7\textwidth]{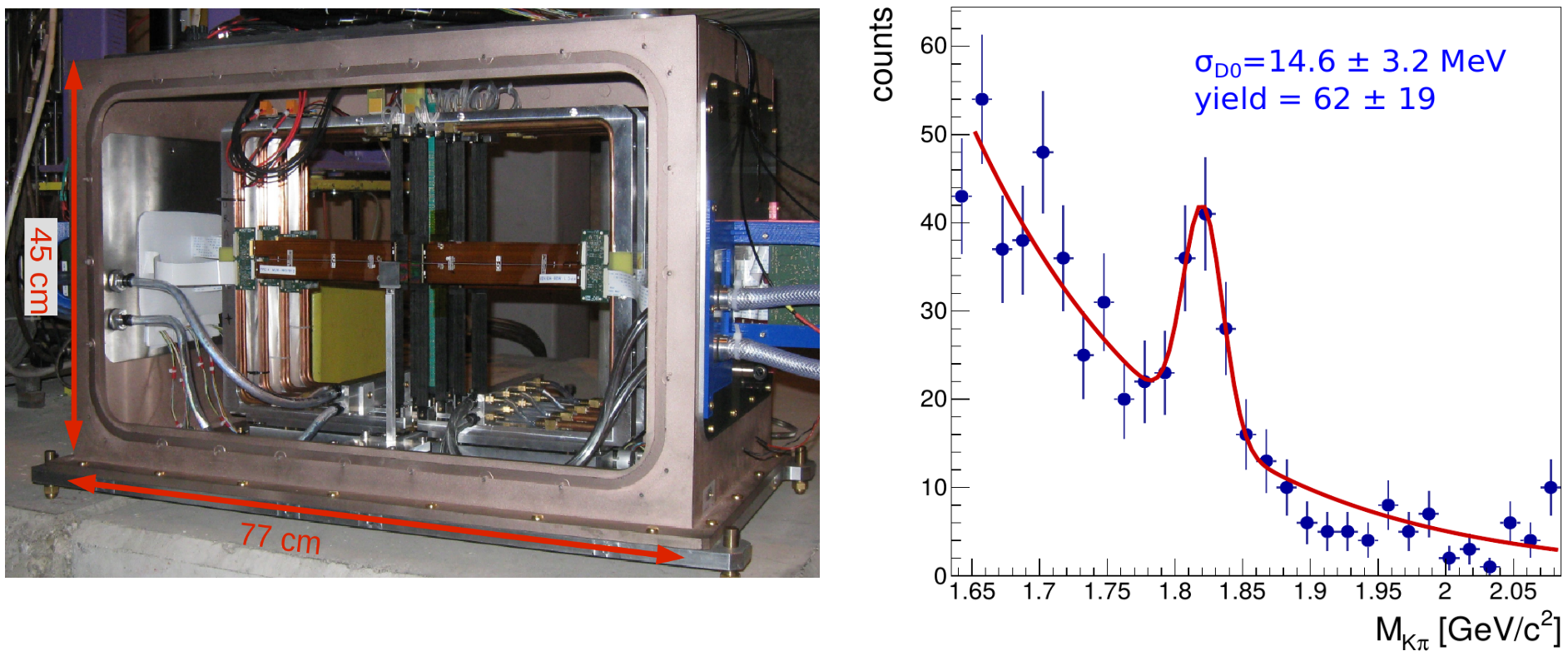}
        \vspace{-0.25cm}
	\caption{\textit{Left:}The SAVD used by NA61/SHINE during the data taking in 2016 and 2017. 
          \textit{Right:} The invariant mass distribution of \Dzero and \Dzerobar candidates in central Pb+Pb 
          collisions at 150\AGeVc after the background suppression cuts. The particle identification capability of 
          NA61/SHINE was not used at this stage of the analysis~\cite{ana1}.
        }
	\label{fig:firstD0}
\end{figure}

The first test of the device was performed in December 2016 during a Pb+Pb test run. 
The test allowed to demonstrate: tracking in a large track multiplicity environment, 
precise primary vertex reconstruction, TPC and SAVD track matching. Furthermore, it allowed to make a first
search for the \Dzero and \Dzerobar signals.
The obtained primary vertex resolution along the beam direction of 30~$\mu$m  
was sufficient to perform the search for the \Dzero and \Dzerobar signals. 
Figure~\ref{fig:firstD0} (\textit{right}) shows the first indication of a \Dzero and \Dzerobar peak obtained 
using the data collected during the Pb+Pb run in 2016.

Successful performance of the SAVD in 2016 led to the decision to also use it during the Xe+La data taking in 2017.
About $5\cdot10^6$ events of central Xe+La collisions at 150\AGeVc were collected in October and November 2017.
During these measurements the thresholds of the MIMOSA-26 sensors were tuned to obtain high hit detection efficiency
which led to significant improvement in the
primary vertex reconstruction precision, namely the spatial resolution of the primary vertices obtained for
Xe+La data is on the level of  $1 ~\mu m$ and $15 ~\mu m$ in the transverse and longitudinal coordinates, respectively.
The distribution of the longitudinal coordinate ($z_{prim}$) of the primary vertex is shown 
in Fig.~\ref{fig:vd-prim-longitu} (\textit{left}) (see Ref.~\cite{ana1} for details) 
The Xe+La data are currently under analysis and are expected to lead to physics results in the coming months.

The SAVD will also be used during three weeks of Pb+Pb data taking in 2018. 
About $1\cdot10^7$ central collisions should be recorded and 2500 \Dzero and \Dzerobar decays can be expected to be reconstructed 
in this data set.

\section{Proposed measurements after LS2}
\label{sec4}


During the Long Shutdown 2 at CERN (2019-2020), a significant modification of the NA61/SHINE spectrometer 
is planned. The upgrade is primarily  motivated by the charm program which requires a tenfold increase of the data taking rate
to about 1~kHz and an increase of the phase-space coverage of the Vertex Detector by a factor of about 2. 
This, in particular, requires construction of a new Vertex Detector (VD), replacement of the TPC read-out electronics, 
implementation of new trigger and data acquisition systems and upgrade of the Projectile Spectator Detector.
Finally, new ToF detectors are planned to be constructed for particle identification at mid-rapidity.
This is mainly motivated by possible future measurements related to the onset of fireball formation.
\begin{figure}[t]
	\centering 
	\includegraphics[width=0.95\textwidth]{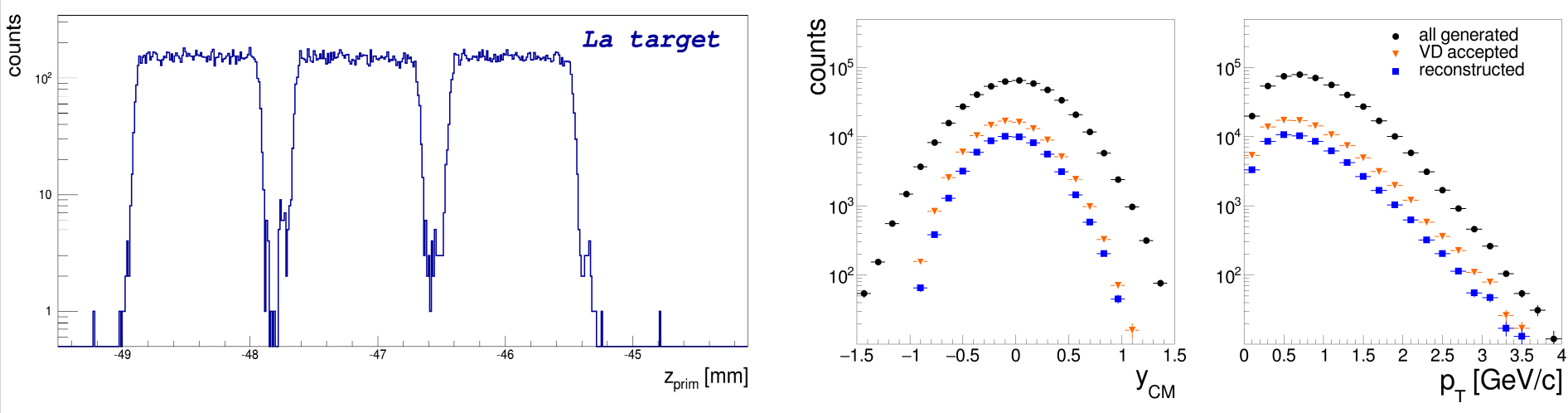} 
        \vspace{-0.3cm}
	\caption{({\it Left:}) Distribution of longitudinal coordinate of the primary vertex $z_{prim}$ for 
          interactions in the La target, which was composed of three 1~mm plates.
          {\it Right:} Rapidity (\textit{left}) and  transverse momentum (\textit{right}) distributions 
	  of \Dzero+ \Dzerobar mesons produced in about 500M inelastic Pb+Pb collisions at 150\AGeVc.   
	  Dots indicate all generated mesons, triangles mesons within the VD acceptance and 
	  squares mesons within the VD acceptance and passing background suppression cuts.} 
	\label{fig:vd-prim-longitu} 
\end{figure}
The detector upgrades are discussed in detail in Ref.~\cite{ana1}.
The data taking plan related to the open charm measurements forsees
measurements of 500M inelastic Pb+Pb collisions at 150\AGeVc in 2022 and 2023.
This data will provide the mean number of 
\ccbar pairs in central Pb+Pb collisions needed to investigate the
mechanism of charm production in this reaction.
Moreover, the data will allow to establish the centrality dependence of \accbar in Pb+Pb collisions
at 150\AGeVc and thus address the question of how the formation of QGP impacts $ \text{J}/\psi $ production.  
Table~\ref{tab:centrality} lists the expected number
of charm mesons in centrality selected Pb+Pb collisions at 150\AGeVc assuming the above mentioned statistics of
minimum bias collisions. The estimate was performed 
assuming that the mean multiplicity of charm hadrons is proportional to the number of collisions and
used yields calculated for central Pb+Pb collisions within the HSD model~\cite{Linnyk:2008hp,TSong}.
Central (0-30\%) Pb+Pb collisions at 40\AGeVc are planned to be recorded in 2024.
This data together with the result for central Pb+Pb collisions at 150\AGeVc will start a long-term
effort to establish the collision energy dependence of \accbar and address the question of how
the onset of deconfinement impacts charm production.
\begin{table}[!htb]
  \caption{		
    Expected number of charm mesons in centrality selected Pb+Pb collisions at 150\AGeVc assuming 500M minimum
    bias events recorded in 2022 and 2023, see text for detail. The mean number of wounded nucleons $\langle W \rangle$ 
    calculated within the Wounded Nucleon Model is also given.
  }
  \label{tab:centrality}
  \centering
  \begin{tabular}{c c c c c c c }
    &   0--10\%  & 10--20\% & 20--30\%  &  30--60\%   &  60--90\%  &  0--90\%   \\ \hline 
    \#(\Dzero+ \Dzerobar)      &    31k    &   20k     &    11k    &  13k        &   1.3k       &  76k      \\
    \#(\Dplus+ \Dminus)        &    19k    &   12k     &    7k    &    8k        &   0.8k     &    46k      \\
    $\langle W \rangle$        &    327    &   226     &   156     &  70         &   11       &  105       \\
  \end{tabular}
\end{table}
The expected high statistics of reconstructed \Dzero and \Dzerobar decays is due to the high event rate 
and the relatively large  efficiencies of open charm detection in the VD. 
The efficiency will be about 13\% (3 times better than for the SAVD) for 
the \Dzero  $\rightarrow \pip + \km$ decay channel and about 9\% 
\footnote{The quoted efficiencies include the geometrical acceptance for 
\Dzero $\rightarrow \pip + \km$ (\Dplus $\rightarrow \pip + \pip + \km$) decays 
and the efficiency of the analysis quality cuts used to reduce the combinatorial background.} 
for \Dplus decaying into $\pip + \pip + \km$. 

Figure~\ref{fig:vd-prim-longitu} (\textit{right})  
shows distributions of \Dzero + \Dzerobar mesons in rapidity  and transverse
momentum for all generated particles (black symbols)  and for particles that passed the acceptance and background
reduction cuts (blue symbols). The presented plots refer to 500M inelastic Pb+Pb collisions at 150\AGeVc. 
Total uncertainty of $ \langle \Dzero \rangle $ and $ \langle \Dzerobar \rangle $ is expected to be about 10\% 
and is dominated by systematic uncertainty.

In summary it is emphasized that only NA61/SHINE is able to measure open charm production in heavy ion collisions 
in full phase space and at the beginning of the next decade. 
The corresponding potential measurements at higher (LHC, RHIC) and lower (FAIR, J-PARC) energies are necessary 
to complement the NA61/SHINE results and establish the collision energy dependence of charm production.

{\bf Acknowledgments:} this work was supported by  the Polish National Center for Science
grants \\ 2014/15/B/ST2/02537 and 2015/18/M/ST2/00125.





\bibliographystyle{elsarticle-num}
\bibliography{<your-bib-database>}



\end{document}